\documentclass[prb,twocolumn,amsmath,floats,showpacs,superscriptaddress]{revtex4}

\usepackage{graphicx}
\usepackage{dcolumn}

\begin{document}
\title{The molecular signature of highly conductive metal-molecule-metal junctions}

\author{O. Tal}
\affiliation{Kamerlingh Onnes Laboratory, Leiden University, P.O.
Box 9504, 2300 RA Leiden, The Netherlands}

\author{M. Kiguchi}
\thanks{Present address: Graduate School of Science, Hokkaido University,
Sapporo, Japan, and JST-PRESTO} \affiliation{Kamerlingh Onnes
Laboratory, Leiden University, P.O. Box 9504, 2300 RA Leiden, The
Netherlands}

\author{W.H.A. Thijssen}
\affiliation{Kamerlingh Onnes Laboratory, Leiden University, P.O.
Box 9504, 2300 RA Leiden, The Netherlands}

\author{D. Djukic}
\affiliation{Kamerlingh Onnes Laboratory, Leiden University, P.O.
Box 9504, 2300 RA Leiden, The Netherlands}

\author{C. Untiedt}
\affiliation{Departamento de Fý´sica Aplicada, Universidad de
Alicante, Campus de San Vicente del Raspeig, E-03690 Alicante,
Spain}

\author{R.\,H.\,M.\ Smit}
\affiliation{Kamerlingh Onnes Laboratory, Leiden University, P.O.
Box 9504, 2300 RA Leiden, The Netherlands}

\author{J.M. van Ruitenbeek}
\affiliation{Kamerlingh Onnes Laboratory, Leiden University, P.O.
Box 9504, 2300 RA Leiden, The Netherlands}

\begin{abstract}

The simplicity of single-molecule junctions based on direct
bonding of a small molecule between two metallic electrodes make
them an ideal system for the study of fundamental questions
related to molecular electronics. Here we study the conductance
properties of six different types of molecules by suspending
individual molecules between Pt electrodes. All the molecular
junctions show a typical conductance of about 1G$_0$ which is
ascribed to the dominant role of the Pt contacts. However, despite
the metallic-like conductivity, the individual molecular signature
is well-expressed by the effect of molecular vibrations in the
inelastic contribution to the conductance.
\end{abstract}

\date{\today}
 \maketitle


\section{introduction}

When forming a conductive molecular bridge between two metallic
electrodes there is a tradeoff between the preservation of the
molecular electronic structure and the realization of highly
conductive molecular junctions. The first quality can be achieved
by molecules that are connected to the electrodes via anchoring
side groups ~\cite{Chen2006} (e.g. Au/benzene
dithiol~\cite{Fujihira2006}) which act as potential barriers that
decouple to a large extent the molecule from both contacts. Such
junctions have a conductance in the tunneling regime (about
10$^{-5}$$-$10$^{-2}$G$_0$ where G$_0$=$2e^{2}/h$ is the
conductance quantum).~\cite{Ulrich2006} The second quality is
accomplished by simple molecules connected directly to the
electrodes (e.g. a Pt/H$_{2}$ junction~\cite{Smit2002}) which
result in a strong molecule-electrode coupling. Such junctions
have a conductance in the (quasi-ballistic) contact regime (about
0.1-1G$_0$)~\cite{Smit2002,Kiguchi2007a,Tal2008,Kiguchi2008} which
is comparable to the conductance of metallic atomic
junctions.~\cite{Agrait2003}

The use of anchoring side groups enables a better a-priori control
over the transport properties of the molecular junction by
chemical synthesis since the structure and properties of the
isolated molecule are preserved to a large extent, and the
orientation of the molecule with respect to the leads is
determined mainly by the position of the anchoring groups on the
molecule. In the case of small molecules that react directly with
the electrodes, the whole molecule serves as an "anchoring group",
thus it is prone to structural and electronic modifications and
its orientation cannot be easily predicted. In some extreme cases
new structures can be formed by molecular decomposition as in the
case of metal atom chains decorated with oxygen
atoms.~\cite{Thijssen2008a}

The direct binding of simple molecules to the electrodes offers a
valuable opportunity to explore some of the central questions
related to electron transport through molecular
junctions.~\cite{Djukic2005,Tal2008,Kiguchi2008} The high
conductivity and the relatively simple electronic and atomic
structure of such junctions permit adapting experimental
techniques that were originally developed for atomic point
contacts (e.g. measurements of shot
noise~\cite{vandenBrom1999,Djukic2006}, conductance
fluctuations~\cite{Ludoph1999,Smit2002} and subgap structure in
superconducting contacts~\cite{Scheer1998,Makk_unpublished2008})
and explore the effect of different manipulations such as junction
stretching and isotope substitution on inelastic
spectroscopy.~\cite{Agrait2002a,Stipe1998,Djukic2005} This variety
of experimental tools extends the number of observed properties
available for research. Moreover simple molecular junctions can be
described to higher accuracy by theoretical calculations due to
the limited size of these physical systems. The simplicity of the
electronic structure may help to validate different
approximations~\cite{Galperin2004,Paulsson2005,Viljas2005,delaVega2006}
that simplify the calculations and provide intuitive
models.~\cite{Paulsson2005,Halbritter2008,Garcia2004} Consequently
the comparison between theory and experiment is more
straightforward.

In view of the role that they may play as reference and model
systems, we present here a comparative study of simple-molecule
junctions based on hydrogen (H$_2$), water (H$_2$O), carbon
monoxide (CO), carbon dioxide (CO$_2$), ethylene (C$_2$H$_2$) and
benzene (C$_6$H$_6$) connected to platinum (Pt) electrodes. We
measured the typical conductance, conductance evolution during
introduction of the target molecule, response to current induced
heating, inelastic spectroscopy and its dependence on junction
stretching. Throughout the paper we focus on the manifestation of
the molecular characteristic properties in the conductance.

\section{Experimental Technique}

The molecular junctions were formed using the mechanically
controllable break junction (MCBJ) technique.~\cite{Muller1992}
Starting with a macroscopic Pt wire (poly-crystalline, 0.1 mm
diameter, 99.99\% purity), a small notch was cut at the middle of
the wire in order to fix a breaking point. The wire was glued on
top of a bending substrate which is mounted in a three-point
bending configuration inside a vacuum chamber that is pumped to a
pressure below 1$\cdot$10$^{-5}$ mbar. After cooling to about
4.5~K, when a cryogenic vacuum is attained, the wire is broken at
the notch by bending of the substrate onto which it has been
fixed. The clean, freshly exposed metallic apexes are then brought
back into contact by slightly relaxing the bending. With the use
of a piezoelectric element the displacement of the two electrodes
can be finely adjusted to form a stable contact of atomic size.
Due to the bendable sample, the contact displacement is about
10$^{3}$ times smaller then the piezo displacement allowing
sub-angstrom control over the contact
separation~\cite{Muller1992,Untiedt2002}. The low-temperature MCBJ
provides: a) clean metal contacts produced under cryogenic vacuum,
b) high mechanical stability of the atomic/molecular junction, c)
precise control over the distance between the electrodes, d) fast
formation and break of atomic/molecular junctions that allows
statistical treatment of measurements done over many different
junctions in a short time.

Two-probe conductance (I/V) measurements versus inter-electrode
separation (conductance traces) are done while repeatedly breaking
the contact using a piezoelectric element (junction pulling speed:
$\sim$400 nm/s). A bias voltage is given by a PC DAQ card (e.g.
NI-PCI-6251, 1.25 Msample/s or NI-PCI-6030E, 100 Ksample/s) and
the current signal is amplified by a current to voltage amplifier
(SR570) and sent to the PC DAQ card. Differential conductance
(dI/dV) measurements were performed on the molecular junction
using a lock-in amplifier (SR830). For this propose the bias
voltage was modulated with a fixed modulation amplitude of 1mV and
a frequency of 7 kHz, while sweeping the dc bias voltage using the
PC DAQ card in the range of +100 to -100 mV and back. The output
AC current signal was amplified by the mentioned
current-to-voltage amplifier before being introduced to the
lock-in amplifier input. The AC signal at the modulation frequency
was then collected by the PC DAQ card.

\section{Introduction of Molecules}

\begin{figure}[b!]
\begin{center}
\includegraphics[width=8.0cm]{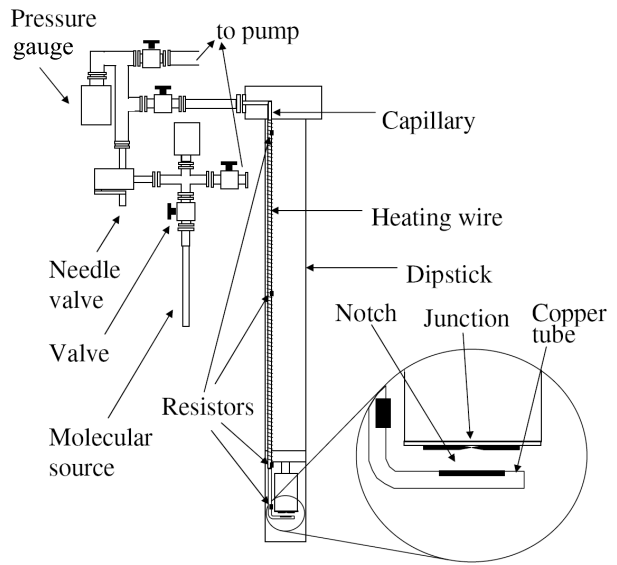}
\end{center}
\caption{Schematic representation of the molecule dozer and its
location within the cryogenic insert.} \label{fig1}
\end{figure}

The introduction of the molecules was done while keeping the
metallic electrodes at a cryogenic temperature. The molecules were
prepared in a different chamber at room temperature. As molecular
sources in the liquid phase we used deionized~\cite{milliq} H$_2$O
and C$_6$H$_6$ (99.9$\%$) placed in a quartz tube. These compounds
were degassed by cycles of freezing, pumping and thawing.
Compounds in the gas phase (H$_2$(99.999$\%$), D$_2$(99.999$\%$;
99.7$\%$ isotope enriched ), CO(99.995$\%$), CO$_2$(99.995$\%$)
and C$_2$H$_2$(99.5$\%$)) were introduced from gas cylinders to a
dozing-chamber that was flushed with the gas and pumped several
times to minimize contaminations. In order to admit the molecules
to the Pt junction at cryogenic temperatures we have used a
molecular dozer (presented schematically in Fig.~\ref{fig1})
containing a stainless steel capillary~\cite{Note_capillary}
located inside the insert tube under cryogenic vacuum with weak
thermal coupling to room temperature and strong thermal coupling
to the liquid-He bath. The dozer can be heated by a heater wire
wrapped around its outer side (or by a thermocoax$^{\copyright}$
heater wire running all along its interior in other versions). A
short removable copper tube with a side nozzle (see Fig.
\ref{fig1} inset) is connected to the end of the main capillary.
This part can be removed when the sample is replaced and it is
designed to emit a molecular jet in a wide angle towards the
junction. The molecular dozer and the gas dozing-chamber were
baked-out before the introduction of molecules. The dozer
temperature is monitored at four different locations along the
capillary by Pt100 resistors~\cite{Quilty2007} and a single
RuO$_2$ resistor~\cite{Fixsen2002} at the cold lower end. This
design provides an efficient and controllable heating of the
capillary up to 360 K and fast cooling below 20K. During heating
the sample temperature is rises by less then 4 K for a few
minutes, as is typical when introducing water. The introduction of
the other molecules involves less heating and the introduction of
H$_2$ or D$_2$ can be done while the capillary maintains its base
temperature.

\begin{figure}[b!]
\begin{center}
\includegraphics[width=8.0cm]{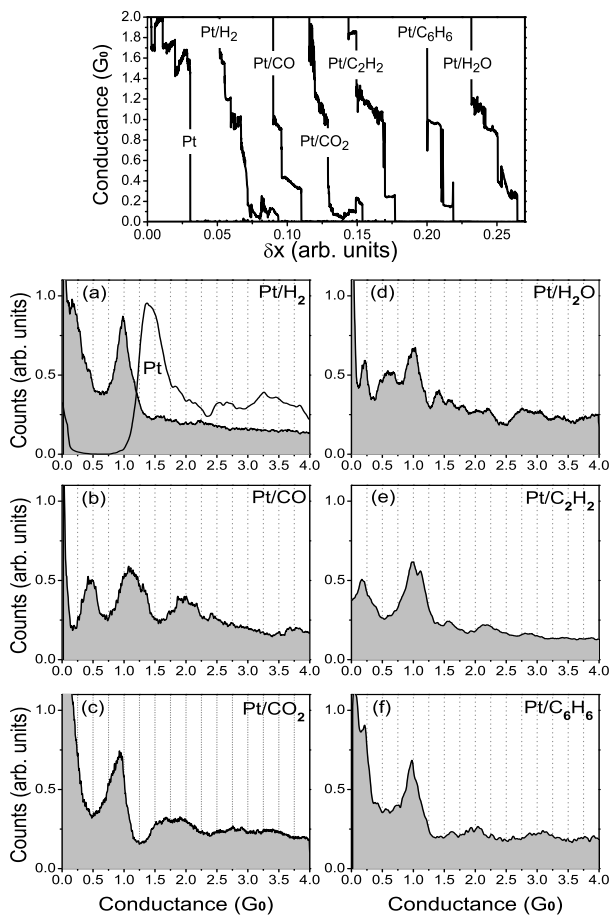}
\end{center}
\caption{Conductance versus displacement traces of Pt junctions
 before (left curve) and after the introduction of each
molecule type. Each trace is composed from 5,000 data points and
located arbitrarily along the displacement axis (top). Conductance
histograms for Pt junctions (normalized to the area under the
curves) before (a, black curve) and after the introduction of
H$_2$(a), CO (b), CO$_2$ (c), H$_2$O (d), C$_2$H$_2$ (e) and
C$_6$H$_6$ (f)(shaded curves). Each conductance histogram is
constructed from more than 1000 conductance traces (bottom).}
\label{fig2}
\end{figure}

While the Pt junction was broken and formed repeatedly, molecules
were introduced to the junction through the molecular dozer.
First, the target molecules were introduced to the cold capillary
via a needle-valve at the top of the capillary (illustrated in
Fig.~\ref{fig1}). Then the temperature of the capillary was
gradually increased until a change in the conductance was detected
(see the following section). The heating power to the capillary
was kept constant for 1 min. Then the heating was stopped and the
capillary nozzle was allowed to cool down via the contact to the
LHe bath. A rough estimation of the molecular deposition rate,
assuming an equilibrium between the frozen target molecules on the
capillary surface and the molecules that were released by heating,
would be 0.05 molecule per minutes per nm$^{2}$ leading to a
sub-monolayer coverage (on average). Note that the mechanical
deformation and the released heat in the process of junction
breaking and formation can assist the introduction of molecule to
the junction. In the case of H$_2$ around 10 $\mu$mol were
introduced to a cold capillary but the dose arriving at the atomic
contact cannot be determined very precisely due to finite vapor
pressure at the base temperature of 4.5 K.

\section{Conductance histograms}

Before the introduction of molecules the formation of a clean Pt
contact is verified by conductance histograms made from at least
1000 conductance traces taken during repeated contact stretching
as presented in Fig.~\ref{fig2}(a) (black curve). The single peak
around 1.5G$_0$ provides a fingerprint of a clean Pt
contact.~\cite{Untiedt2007} The introduction of the target
molecule is signaled by the suppression of the typical peak for Pt
and the appearance of a new distribution of peaks
(Fig.~\ref{fig2}, filled curves)~\cite{Smit2002,note_control}. The
peak locations for each of the molecular junction is given in
Table 1. Among the six junctions Pt/H$_2$O and Pt/CO have a
somewhat more distinct character: the conductance histogram
for Pt/H$_2$O junctions is characterized by several relatively low
peaks, which implies a variety of stable junction configurations.
In some cases instead of the 1.0G$_0$ peak, two peaks at 0.90 and
1.10G$_0$ can be detected. Pt/CO has a unique peak at around
0.5G$_0$ that appears in about 45\% of the cases. Interestingly,
calculations~\cite{Strange2006,Ferrer2008unpublishedV2} cannot
reproduce the 1G$_0$ peak in the conductance histograms of Pt/CO
while the 0.5G$_0$ is successfully obtained and is attributed
to an asymmetric configuration where each atom (C and O) is
attached to a different electrode.

A peak around 0.2G$_0$ appears in some of the histogram
measurements for Pt/H$_2$, Pt/H$_2$O, Pt/C$_2$H$_2$ and
Pt/C$_6$H$_6$. For Pt/H$_2$ it is attributed to molecular
junctions that involves molecules connected to a chain of Pt
atoms~\cite{Kiguchi2007b}, while for Pt/C$_6$H$_6$ it is was
argued to be associated with the conductance of the stretched
molecular junction.~\cite{Kiguchi2008} The origin of the 0.2G$_0$
conductance peak in the histograms of Pt/H$_2$O and Pt/C$_2$H$_2$
has not been studied.

All the conductance histograms in Fig.~\ref{fig2} reveal a peak at
around 1G$_0$ while its fine location and shape slightly varies
between different molecular junctions. The appearance of this
common peak is rather surprising specifically for Pt/H$_2$ and
Pt/H$_2$O for which it has been demonstrated that the conductance
at the 1G$_0$ peak in the histogram is carried dominantly by a
single conductance channel.~\cite{Smit2002,Tal2008} According to
the Landauer equation G=$\sum_{i}$G$_0$T$_{i}$ where T$_{i}$ is
the i$^{th}$ transmission probability for an electron to cross the
junction. For a single conductance channel, G=G$_0$T, thus the
channel is fully open (T=1) at G=1G$_0$. In a simple picture of a
molecule with discrete electronic levels located between two metal
electrodes, the Fermi level of the source and drain electrodes need to be
aligned with the center of a molecular level to have the maximum
transmission probability ("full resonance" case). In an
alternative picture, electron-electron correlations can lead to a
resonance at the Fermi level under specific
conditions.~\cite{Chiappe2005}

Bearing these pictures in mind, one could argue that it is
remarkable to have the special conditions for such a perfect
resonance for different single-channel molecular junctions. In the
case of molecular junctions with several dominant channels, having
the same total conductance of 1G$_0$ for several different
junctions is surprising as well when considering that each
junction has its own individual channel composition. However, as
described by Ferrer \textit{et al.}~\cite{Ferrer2008unpublishedV2}
a strong hybridization of the molecular levels with the d-band of
the metallic electrodes leads to a transmission probability of
about 1 for a wide energy range around the Fermi level even when
the Fermi energy is not aligned with a molecular
level. Thus, the dominant role of the metallic electrode leads to
a common peak around 1G$_0$ for many molecules, where the
individual signature of the molecules is limited to the fine
structure of the conductance histogram.

Note that an alternative scenario of a metal-to-metal contact
decorated with molecules, or accompanied with a secondary
conductance channel via a metal-molecule-metal bridge is unlikely.
The main conductance peak which is found around 1G$_0$ for the
molecular junctions, lies much lower then the conductance of a
Pt-Pt contact (around 1.5G$_0$). Tuning the distance between Pt
electrodes to achieve 1G$_0$ conductance is almost impossible - in
this case the conductance either jumps above 1.4G$_0$ or to
tunneling conductance below 0.1G$_0$~\cite{Untiedt2007}. In
general terms, the size of a measured inelastic contribution to
the current due to molecular vibrations is 1-10$\%$. When the
current path through the molecule takes only a small fraction of
the total current ($<$10$\%$), as expected for the mentioned
scenario, an inelastic contribution of 1-10$\%$ to the total
current due to that channel would become unphysically large which
makes this scenario less likely.

\begin{table}[t!]
\begin{center}
\begin{tabular}{|p{1.0cm} p{1.0cm} p{1.0cm}p{1.0cm} p{1.0cm} p{1.0cm}|}
 \hline
  H$_2$ & CO & CO$_2$ & H$_2$O & C$_2$H$_2$ & C$_6$H$_6$ \\ \hline
  $0.20$ & $0.46$ & $0.94$ & $0.20$ & $0.22$ & $0.21$ \\
  $1.00$ & $1.10$ & $ $ & $0.60$ & $1.04$ & $0.98$ \\
  $ $ & $ $ & $ $ & $1.00$ & $ $ & $ $ \\ \hline
\end{tabular}
\end{center}
\caption{\small{Conductance (in units of G$_0$ and $\pm$0.05G$_0$
uncertainty) of the main peaks in the conductance histograms of
the studied molecular junctions.}} \label{table}
\end{table}

\section{the evolution of conductance histograms : from metallic contact to molecular junction}

\begin{figure}[t!]
\begin{center}
\includegraphics[width=8.0cm]{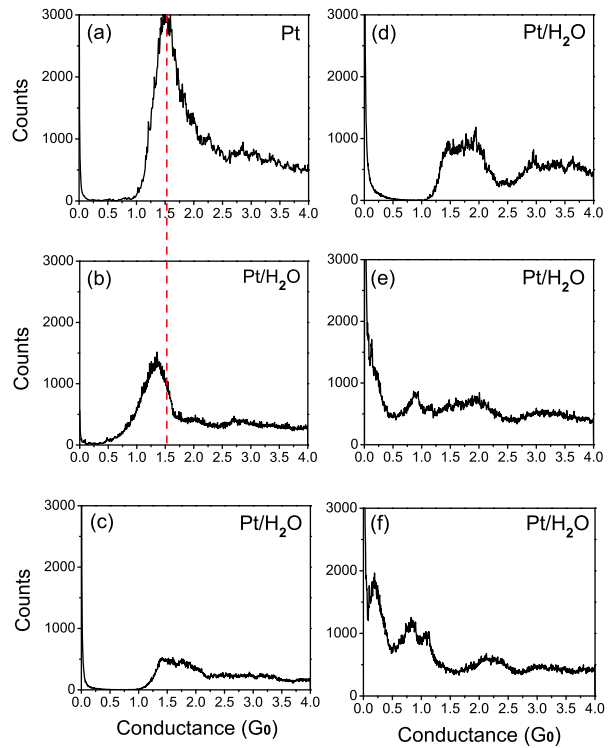}
\end{center}
\caption{(Color online) A series of sequential conductance
histograms (i.e. each consequential histogram is restarted and
does not contain data from the former one) for a virgin Pt
junction (a) and a Pt junction after the introduction of H$_2$O
(b-f). Each conductance histogram is constructed from 1000
conductance traces and takes about 6 min.} \label{fig3}
\end{figure}

The conductance histograms in Fig.~\ref{fig2} were taken at the end of the
process of introducing the target molecules. When the molecules are
first introduced to the metallic junction
the conductance histogram is observed to evolve through a series
of steps that were found to be similar for H$_2$O, CO, CO$_2$, and
C$_6$H$_6$. The introduction of C$_2$H$_2$ was not studied in this
aspect and the evolution of conductance histograms for Pt/H$_2$
junctions is too fast for collecting reliable
statistical data related to the junction evolution. Here we focus
on the evolution of sequential conductance histograms while H$_2$O
molecules are admitted to a Pt junction as presented in
Fig.~\ref{fig3}. A conductance histogram of a virgin Pt junction
is first taken (see Fig.~\ref{fig3} (a)). Then the target
molecules are introduced to the cold capillary followed by gradual
heating of the capillary. When the temperature at the outer side
of the capillary near the sample reaches ~120K (the thermocoax
temperature could be higher), the main peak in the histogram
shifts to lower conductance (in this case from 1.5G$_0$ to
1.4G$_0$, as seen in Fig.~\ref{fig3} (b)) while its amplitude is
suppressed (Fig.~\ref{fig3} (b) and (c)). Next, contributions
around 1.8G$_0$ become more pronounced as demonstrated in
Fig.~\ref{fig3} (d). Finally the feature around 1.8G$_0$ is
suppressed and a conductance histogram typical for the final
molecular junction has evolved (Fig.~\ref{fig3} (e and f)).

The shift of the main peak in Fig.~\ref{fig3} (a) to lower
conductance can stem from different origins: (1) stimulated
formation of Pt chains by the presence of molecules around the
metallic contact can lead to lower conductance around
1.2-1.4G$_0$.~\cite{Smit2003,Garsia-Suarez2005} Indeed a shift of
the Pt peak to $~$1.2G$_0$ is sometimes observed; (2) a lower
conductance can arise from a weakening of the Pt-Pt bonds at the
Pt junction by the adsorption of molecules near the junction (due
to electrostatic effects, charge transfer, or other
perturbations). These are two mutually excluding possibilities
since the first scenario involves strengthening of the Pt-Pt bonds
leading to chain formation when the contacts are pulled apart,
while the second scenario is based on bond weakening.

When the shifted peak is suppressed, the conductance around
1.8G$_0$ becomes more pronounced. This value agrees with the
conductance through a single bond between two Pt
electrodes~\cite{Smit2003,Nielsen2003} or more specifically,
between two pyramidal Pt electrodes.~\cite{Garsia-Suarez2005} The
number of counts below 1G$_0$ remains low which indicates that
there are no molecular bridges being formed. This evolution step
can be interpreted as a suppression of chain formation (which
would give the 1.4-1.5G$_0$ peak) by the presence of molecules
(i.e. weakening of the Pt-Pt bond). Thus the scenario of adsorbate
induced weakening of the Pt-Pt bonds at the junction can explain
both the shift of the 1.5G$_0$ peak to lower conductance values
and the increased probability for conductance at 1.8G$_0$ at the
expense of the former peak. The sequence of events is in agreement
with this scenario as well since moderate weakening of the Pt-Pt
bond is expected to reduce conductivity while further weakening
will lead eventually to bond breaking and suppression of chain
formation.

\section{the effect of current-induced heating on conductance histograms}

As demonstrated in Fig.~\ref{fig3} conductance histograms form an
efficient tool in probing the presence of conducting molecules in
the junction (and probably even the presence of molecules in the
vicinity of the metallic junction). However, for most of the cases
examined discrimination between the different molecules just from
the conductance histogram is not trivial, as can be seen in
Fig.~\ref{fig2}. Thus, additional tools are required in order to
distinguish between them.

The evolution of the conductance histograms following application
of high enough bias voltage (typically 1 V for 1 min) across the
junction varies for different molecular junctions. For Pt/H$_2$
junction a typical histogram for clean Pt is recovered by application of
such a high voltage. However, upon reducing the voltage below 300
mV the typical conductance histogram of Pt/H$_2$ reappears. This
behavior can be understood as the response to current-induced
heating at the junction.~\cite{Todorov1998,Segal2002,Alavi2003} As
a consequence of local heating of the junction the hydrogen
molecules diffuse away from the contact. However due to the finite
diffusion rate of the light hydrogen at ~4.5 K a Pt/H$_2$ junction
is recovered once the heating has stopped. For CO, CO$_2$, and
C$_6$H$_6$ the application of a high bias results in the recovery
of a clean Pt histogram that remains stable even after
reduction of the applied voltage. We ascribe the latter observation to
the negligible diffusion rates for these molecules at $~$4.5 K.
Finally, high voltage usually does not affect the
conductance histogram of Pt/H$_2$O junctions implying a relatively
stable binding of the molecule to the electrodes. In some rare
cases a Pt histogram can be obtained provided that the high
voltage was applied a few seconds after the first indication for
H$_2$O in the junction. The difference in response following the
application of a high bias voltage provides a quick method for
obtaining an indication on the nature of the molecular junction.

\section{inelastic spectroscopy}

\begin{figure}[t!]
\begin{center}
\includegraphics[width=8.0cm]{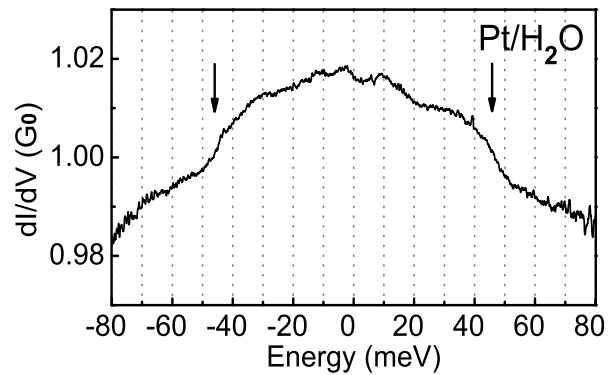}
\end{center}
\caption{Differential conductance (dI/dV) as a function of applied
voltage across a Pt/H$_2$O junction. The steps marked by the
arrows are due to inelastic scattering by vibration modes in the
molecule.} \label{fig4}
\end{figure}

For inelastic spectroscopy the differential conductance (dI/dV) is
measured as a function of the applied voltage across a molecular
junction. Fig.~\ref{fig4} provides an example of a measured
spectrum. When the voltage difference between the metal electrodes
reaches a threshold given by a vibration mode energy
($\hbar\omega$) the electrons have enough energy to excite the
molecular vibration mode.~\cite{Stipe1998,Djukic2005} At this
voltage (eV=$\hbar\omega$) a step in the differential conductance
appears due to the change in the transmission probability for
electrons that interact with the molecular
vibration~\cite{Tal2008}(e.g. at 46 meV in Fig.~\ref{fig4}). The
interpretation of the step in the dI/dV signal in terms changes
induced by a molecular vibration mode has been confirmed by
measurements of isotopes~\cite{Djukic2005,Kiguchi2008} D$_2$, HD
and $^{13}$C$_6$H$_6$.

\begin{figure}[t!]
\begin{center}
\includegraphics[width=8.0cm]{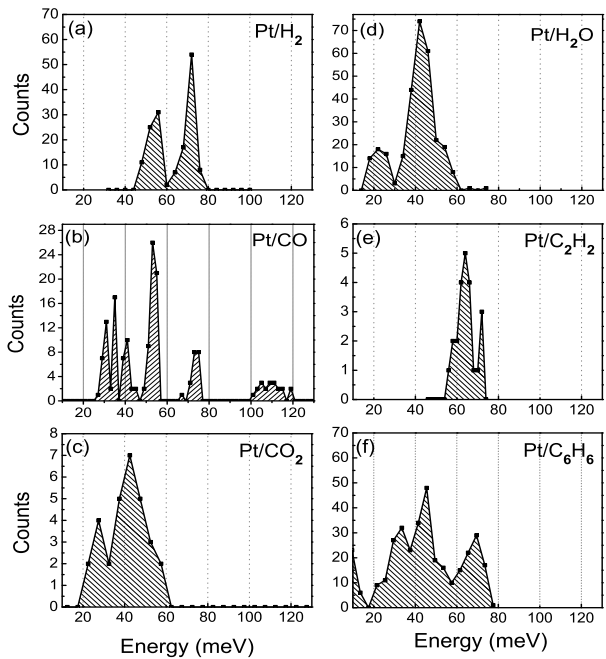}
\end{center}
\caption{Distribution of vibration mode energies observed for
H$_2$(a), CO (b), CO$_2$ (c), H$_2$O (d), C$_2$H$_2$ (e) and
C$_6$H$_6$ (f) between Pt electrodes. The data was collected for
junctions with zero-bias conductance below 1.1G$_0$ (for
C$_6$H$_6$ this range is wider then appears in a previous
report~\cite{Kiguchi2008} where the data was collected for
zero-bias conductance below 0.4G$_0$). Note that for Pt/CO and
Pt/C$_2$H$_2$ junctions only a limited set of data was collected.}
\label{fig5}
\end{figure}

For each type of molecular junction we have collected many $dI/dV$
spectra where each measurement is performed on a newly formed
junction. Fig.~\ref{fig5} presents the number of times that a
vibration mode with certain energy was detected for each molecular
junction. The distribution of the typical vibration mode energies
reveals an individual signature for a specific molecular junction.
The following two cases demonstrate the implications of such a
chemical recognition.

On a few occasions, after several days of measurements taken on
Pt/C$_6$H$_6$ and also Pt/CO junctions, unexpected features in the
conductance histograms suggested a possible presence of H$_2$O in
the Pt junctions. The different response to current induced
heating experiments (see section VI) supported this
interpretation. However, the different vibration energy
distributions for these junctions gave us much stronger indication
for the presence of H$_2$O as a contamination in the studied
junctions. Thus the application of vibration mode distribution as
a chemical fingerprint of the junction allows us to deduce that
substitution of the target molecules by contaminating molecules
with higher affinity can take place even at low temperatures.

Chemical recognition is also important in cases where the
conductance histogram of the target molecule is similar to that
for junction based on a derivative of the target molecule as in
the case of Pt/CO and Pt/CO$_2$ junctions. The additional peak
around 0.5 G$_0$ for the Pt/CO junction does not appear in all
cases, leading to similarity between the conductance histograms
for the two junctions. However, the large difference between the
vibration energy distributions for these junctions allows us to
identify the presence of each molecule in the junction.

\section{stretching dependence in inelastic spectroscopy}

\begin{figure}[t!]
\begin{center}
\includegraphics[width=8.0cm]{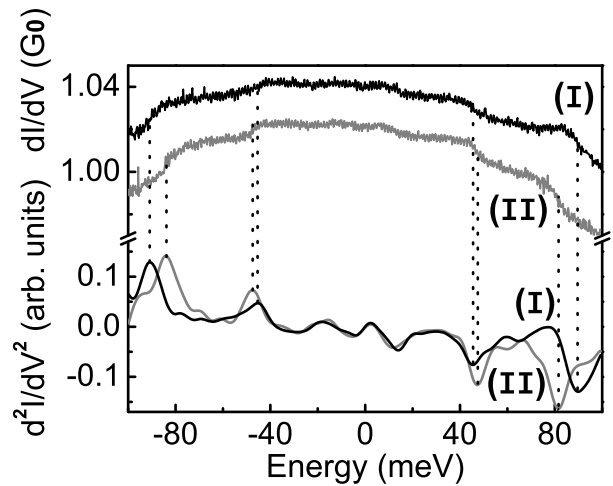}
\end{center}
\caption{dI/dV (upper two curves) and corresponding numerical
derivatives d$^{2}$I/dV$^{2}$ (lower two curves) as a function of
the energy. Both measurements I) and II) are obtained on the same
junction, except that the junction is stretched before the
measurement II).} \label{fig6}
\end{figure}

The MCBJ technique provides sub-atomic control on the distance
between the electrodes while the mechanical stability of the
junction is maintained. When using MCBJ in combination with simple
molecules the dependence of vibration energies on junction
stretching can provide information on the symmetry of the
vibration mode.

Figure~\ref{fig6} shows differential conductance curves that were
taken on the same Pt/D$_{2}$ junction before (I) and after (II)
stretching the junction. Two steps can be seen in each curve,
which correspond to two different vibration modes. Due to our
limited energy window for inelastic spectroscopy measurements
(above ~120 mV the Pt/H$_{2}$ junction is unstable) we use D$_{2}$
to shift  these vibration energies down towards the center of our
measurement window as a result of the twice larger
mass.~\cite{Djukic2005} Focusing on the differences between curves
(I) and (II), one observes an increase in the lower energy mode
and a decrease in the energy of the higher mode due to junction
stretching. This is more clearly seen in the shifts of the
peaks/deeps in the conductance derivative (dI$^{2}$/d$^{2}$V).

\begin{figure}[t!]
\begin{center}
\includegraphics[width=8.0cm]{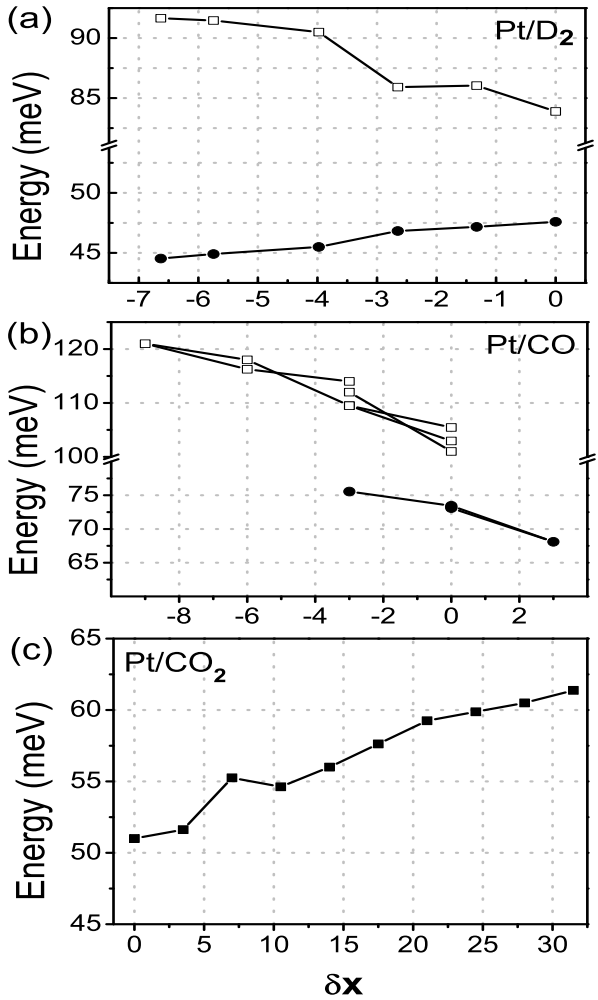}
\end{center}
\caption{Vibration energy as a function of the stretching for
Pt/D$_2$ (a), Pt/CO (b) and Pt/CO$_2$ (c) junctions. The
stretching distance is measured with respect to the starting point
denoted as zero. The unit for the displacement ($\delta$x) axis is
about 0.1$\pm$0.03~{\AA}.} \label{fig7}
\end{figure}

Figure~\ref{fig7} (a) presents sequential vibration energy shifts
due to several stages of stretching. The energy increase of the
lower mode (filled circles) can be explained by the response of a
transverse mode to stretching, in analogy to a guitar string which
gives a higher pitch upon stretching due to an increase in the
restoring force. The reduction in the higher mode energy (hollow
squares) can be ascribe to the effect of stretching on the
longitudinal mode, where the electrode-molecule bond is elongated
and weakened resulting in a frequency reduction. This response to
stretching agrees closely to the results from Density Functional
Theory calculations.~\cite{Djukic2005} Note that the latter effect
of the weakening of the electrode-molecule bond is also relevant
in the case of transverse mode, where following an energy increase
due to stretching, further stretching may lead to reduction in the
vibration energy. Thus observing an energy reduction by junction
stretching is not sufficient for identifying the mode orientation
and further tests and comparison with calculations are required.

In Fig.~\ref{fig7} (b) the vibration energies for two modes of a
Pt/CO junction are reduced by stretching, while the other modes
observed in Fig.~\ref{fig5} (b) change very little by stretching.
We have not succeeded in obtaining a satisfactory interpretation
of the results for Pt/CO in terms of model
calculations.~\cite{Thijssen_unpublished}

Fig.~\ref{fig7} (c)
reveals an increase in the energy for the higher vibration mode of
a Pt/CO$_{2}$ junction. Thus we conclude that this is a transverse
mode. No stretching dependence was found in the case of H$_2$O and
C$_6$H$_6$ while this property has not been studied for
Pt/C$_{2}$H$_{2}$. Interestingly, up to now stretching dependence
was found only in linear molecules.

\section{Summery and concluding remarks}

Measurements of conductance histograms on molecular junctions for
Pt/H$_2$, Pt/H$_2$O, Pt/CO, Pt/CO$_2$, Pt/C$_2$H$_2$ and
Pt/C$_6$H$_6$ show  in all cases a main peak around 1G$_0$ that
indicates the most probable conductance for these junctions. This
general behavior is ascribed to the dominant role of the Pt
metallic electrode in the bonding to the molecules. Note that this
effect is not unique to Pt electrodes and was observed in other
cases as well.~\cite{Untiedt2004,Thijssen2007_thesis} The
signature of the molecule is expressed in the fine structure of
the histograms such as the exact location of the main peak around
1G$_0$ and the peak shape, and in many cases additional peaks are
observed. The evolution of the conductance histograms following
the introduction of the molecules was found to be similar for the
different junctions probably because the process is governed by
the metal electrodes.

The molecular nature emerges in the response of the junction to
current induced heating. The junction response can be classified
in three scenarios: (i) recovery of a clean Pt junction followed
by reconstruction of the molecular junction shortly after the
heating (Pt/H$_2$), (ii) recovery of a clean and stable Pt
junction (Pt/CO, Pt/CO$_2$, Pt/C$_2$H$_2$, Pt/C$_6$H$_6$),(iii) no
significant effect (Pt/H$_2$O). The different behavior is
attributed to different molecule diffusion rates at low
temperatures and different bonding strength.

Finally, inelastic spectroscopy reveals the characteristic
vibration modes of each molecule junction, while the effects of
stretching and isotope substitution on the vibration energy allow
additional classification of the vibration modes. The distribution
of vibration modes given by inelastic spectroscopy is individual
for each molecular junction, thus the chemical signature of the
suspended molecules is clearly preserved.

\section{Acknowledgments}

This work is part of the research program of the ''Stichting
FOM,'' which is financially supported by NWO. OT is greatly
acknowledges the support by IUVSTA-Welch foundation and CU
acknowledges support through grants MAT2007-65487 and "Consolider"
(CSD2007-0010) of the Spanish MEC.


\begin{thebibliography}{37}
\bibitem[{Not()}]{Note_capillary}
\bibinfo{note}{A different version is composed of two sections: an upper
  Stainless steel tube for an efficient thermal isolation from room temperature
  and a lower copper tube for homogeneous and fast heating/cooling.}
\bibitem[{not()}]{note_control}
\bibinfo{note}{In control experiments no indication for contaminations were
  found when the capillary was heated to the relevant temperature without
  introducing the target molecules.}
\end{thebibliography}
\end{document}